\documentclass[twocolumn,showpacs,preprintnumbers,amssymb]{revtex4}

\def\wh{\widehat}
\usepackage{latexsym}
\usepackage{graphicx,epsf, epsfig, amssymb}
\usepackage{bm}
\usepackage{longtable}

\def\be{\begin{equation}}
\def\ee{\end{equation}}
\def\beq{\begin{eqnarray}}
\def\eeq{\end{eqnarray}}

\begin{document}

\title{
Neutron star tidal disruption in mixed binaries:\\
the imprint of the equation of state }

\author{V.~Ferrari$^1$, L.~Gualtieri$^1$, F.~Pannarale$^{1,2}$}
\address{$^1$Dipartimento di Fisica, ``Sapienza'' Universit\`a di Roma
  \& Sezione INFN Roma1, Piazzale Aldo  Moro 5, 00185, Roma, Italy\\
  $^2$Max-Planck-Institut f\"ur Gravitationsphysik,
  Albert-Einstein-Institut, Potsdam-Golm, Germany}

\begin{abstract} 
  We study the tidal disruption of neutron stars in black hole-neutron
  star coalescing binaries.  We calculate the critical orbital
  separation at which the star is disrupted by the black hole tidal
  field for several equations of state describing the matter inside
  the neutron star, and for a large set of the binary parameters.
  When the disruption occurs before the star reaches the innermost
  stable circular orbit, the gravitational wave signal emitted by the
  system is expected to exhibit a cutoff frequency $\nu_{GWtide}$,
  which is a distinctive feature of the waveform.  We evaluate
  $\nu_{GWtide}$ and show that, if this frequency will be found in a
  detected gravitational wave, it will allow one to determine the neutron
  star radius with an error of a few percent, providing valuable
  information on the behaviour of matter in the stellar core.
\end{abstract}

\maketitle

\section{Introduction}\label{intro}
The coalescence of neutron star-neutron star (NS-NS) and black hole-neutron star
(BH-NS) binaries is one of the most promising sources of gravitational
waves to be detected by ground based gravitational wave detectors like Virgo and LIGO 
\cite{virgoligo}.  
These detectors have now reached the planned sensitivity and they will evolve 
toward a second generation, the advanced (Virgo and LIGO) detectors, 
with a sensitivity enhanced by an order of magnitude. 
Furthermore, a design study for an even more sensitive third generation 
detector, ET (Einstein Telescope), is in progress \cite{et}.

In a recent study, based on a population synthesis approach
\cite{Sadowski_etal}, the formation and the evolution of compact binary
systems has been followed from the onset of star formation, both in the
Galactic field, where their massive binary progenitors evolve in isolation,
and in dense clusters, where they can form at high rates due to dynamical
interactions.  The authors estimate that advanced LIGO should detect the
merger of NS-NS binaries at a rate of $\sim 15$ events per year and the merger
of BH-NS binaries at a rate of $\sim1$ per year.  Similar estimates hold for
the advanced version of Virgo.  Thus, it is reasonable to expect that in a
near future we may be able to detect the gravitational wave (GW)  signals 
emitted by these sources and study their features.

A further reason to be interested in these coalescing binaries 
is that they have been proposed 
as providing the engine for short gamma-ray bursts (SGRBs); the detection of a 
gravitational wave signal emitted by one of these systems in coincidence with a SGRB would 
validate this model, thus clarifying one of the most interesting open issues in 
astrophysics.

In this paper we study the disruption of a neutron star in a BH-NS coalescing 
binary, to envisage a method to extract information on the equation of state 
(EOS) of matter in the NS interior from the detection of the gravitational wave 
signal emitted in the process.  We use the \emph{affine model} approach 
\cite{CL85},\cite{LM85},\cite{WL2000},\cite{LS76}, 
which treats the NS as an extended object, 
which responds to its self-gravity, to its internal pressure forces and 
to the relativistic BH tidal field, under the constraint that its shape is 
always that of an ellipsoid.  In the original formulation of this model, 
the internal structure of the star was treated at a Newtonian level and 
the EOS was assumed to be polytropic.  In \cite{tidaldisruption1} 
(to be referred to as Paper I  hereafter) we have improved this approach 
by introducing a general relativistic description of the stellar structure, 
and generalizing the relevant equations to include more general equations of 
state.  This improved approach has been applied to study quasi-equilibrium 
sequences of BH-NS binaries, and to determine the critical orbital separation at 
which the star is torn apart by the black hole tidal field.

When the NS is disrupted before reaching the Innermost Stable Circular 
Orbit (ISCO), the emitted GW signal is expected to 
change abruptly and its amplitude is expected to decrease sharply; such 
signal should exhibit a cutoff frequency $\nu_{GWtide}$, corresponding
to the orbit at which the disruption occurs, i.e.  
$\nu_{GWtide}= 2\nu_{orb~tide}$. This prediction is confirmed by 
numerical simulations where the coalescence of a black hole and a 
neutron star and the tidal disruption of the star have been studied i) 
in the framework of Newtonian gravity \cite{LeeKluzniak1999}, 
\cite{Lee2001}, and ii) in full general relativity 
\cite{Zachariasetal}, \cite{rantsiouetal2008}.  

The frequency cutoff is a distinctive feature of the waveform emitted by 
a BH-NS coalescing binary and indicates the disruption of the star.  
This was first pointed out in \cite{vallisneri}, where a relation between 
$\nu_{GWtide}$ and the stellar radius was derived, using 
the formulation of the affine 
model appearing in \cite{shibata1996} and describing the inspiralling of an 
incompressible, homogeneous, Newtonian ellipsoid moving on circular 
orbits around a rotating black hole. Hints on the role of the 
compressibility were derived in \cite{vallisneri} using some 
results of \cite{WL2000}, where the same process was studied for a 
polytropic, Newtonian star.

In this paper we study the relation between the cutoff frequency and 
the neutron star EOS. Using the same approach developed in Paper I, 
we explicitly compute for which values of the masses of the binary components 
and of the black hole angular momentum the star is tidally disrupted before 
reaching the ISCO, for a variety of realistic equations of state.  
Knowing the disruption distance $r_{tide}$, we evaluate 
$\nu_{GWtide}$ and show how a measure of this quantity from 
a detected  GW signal may be used to infer interesting information 
on the equation of state of matter in the neutron star interior.

It is known that the fully relativistic numerical study of the last 
phases of BH-NS binary coalescence is a quite difficult task, 
especially for large values of the mass ratio $q=M_{BH}/M_{NS}$; this 
is due to a number of reasons, which include the lack of symmetry,
the difficulty in evolving the NS while handling the BH singularity, 
and the prohibitively high 
computational costs required to span the parameter space 
(mass ratio, EOS, spins etc.). For these reasons, the literature 
on the subject is limited to a restricted number of studies, 
most of which are in the framework of Newtonian gravity.  
BH-NS coalescence in general relativity has been studied in 
the following papers (see \cite{Faber2009} for a recent 
review on the subject): non spinning black hole and 
$M_{BH} >> M_{NS}$ in \cite{faberetal2006a}; $q= 10$ 
in \cite{faberetal2006b}; $q=1,3,5$ in\cite{Zachariasetal}; 
and $M_{BH}= 3.2 M_\odot$, $M_{BH}= 4 M_\odot$ and 
$M_{NS}=1.4 M_\odot$ in \cite{shibatauryu2007}.  
In \cite{tanuguchietal2007}, quasi-equilibrium sequences of 
black hole-neutron star binaries have been studied in general 
relativity for $q= 1,2,3,5,10$.  The case of a rotating black 
hole has been considered in \cite{Zachariasetal}, 
where the coalescence has been studied for three values of 
the black hole angular momentum ($a/M_{BH}=-0.5,0,0.75$) and 
for $q=3$, whereas in \cite{rantsiouetal2008} the case 
$q=10$ has been investigated.  It should be stressed that 
in {\em all} these studies the neutron star is modeled using 
a polytropic equation of state.

The improved affine model approach which we use in our study is an
approximate method, but it has the advantage of allowing one to
explore a large region of the parameter space, including large values
of the mass ratio $q=M_{BH}/M_{NS}$, and to use modern equations of
state to model matter in the neutron star interior, at a much lower
computational cost than fully relativistic simulations.

With respect to ref. \cite{vallisneri} our study introduces 
several novelties: we describe the internal structure of the star 
using general relativity, we explicitly take into account  the stellar 
compressibility and describe the matter inside the star using 
realistic equations of state; in addition, in \cite{vallisneri} 
$\nu_{GWtide}$ was 
provided by the Kerr geometry in the approximation of a
point mass NS following geodesics in this geometry, while we determine
this quantity using the 2.5 Post Newtonian equations of motion for the
orbital dynamics of a binary system.

The plan of the paper is the following. 
In Section \ref{affinemodel} we briefly describe the affine 
approach; in Section \ref{nutide} we discuss how to  determine 
the cutoff frequency $\nu_{GWtide}$. The equations of state we employ
are briefly described  in Section \ref{eos};
in Section \ref{errorbar} we discuss the errors which affect the measure of 
the binary parameters from a detected GW signal, and how they propagate
and affect the evaluation of $\nu_{GWtide}$ and of the neutron star radius.
The results of our study are reported in Section \ref{results} and 
in Section \ref{concl} we draw our conclusions.
\section{The improved affine model equations}\label{affinemodel}
The equations of the improved affine model approach, which we need in
order to determine the radial distance $r_{tide}$ at which
the neutron star is torn apart by the tidal interaction with the black
hole, are described in detail in Paper I. Here we will summarise the
assumptions underlying this approach and we shall write only the
equations to be solved.
\subsubsection{Relevant assumptions and equations}
We consider a star in the tidal field of a Kerr BH whose center of
mass follows equatorial, circular orbits; while moving, the star
maintains an ellipsoidal shape; more precisely it is a Riemann-S type
ellipsoid, i.e. its spin and vorticity are parallel and their ratio is
constant (see \cite{EFE}). The NS equilibrium structure is determined
using the stellar structure equations of General Relativity, while
dynamical behaviour is governed by Newtonian hydrodynamics improved by
the use of an effective relativistic self-gravity potential. The
equations for the NS deformations are written in the principal frame,
i.e. the frame associated with the principal axes of the stellar
ellipsoid. Tidal effects on the orbital motion are neglected, as well
as the perturbation that the star induces on the BH.

We study the evolution of the system in the quasi-equilibrium
approximation, i.e. we neglect all time derivatives appearing in the
ordinary differential equations of the model. Physically this means
that we assume that the NS follows a quasi-equilibrium sequence during
the BH-NS coalescence; according to the model, moreover, the
circulation $\mathcal{C}$ of the fluid along this sequence is
constant. In particular, we set $\mathcal{C}=0$, that is, we consider
the NS fluid to be irrotational.  In the principal frame, the fluid
variables of the improved affine model are five: the three principal
axes of the stellar ellipsoid $a_1,a_2$ and $a_3$, the angular
frequency of the internal fluid motion $\Lambda$ and the star spin
$\Omega$. The axes $a_1$ and $a_2$ both belong to the orbital plane,
while $a_3$ is perpendicular to it; $a_1$ indicates the axis that lies
along the binary orbital radius. The quasi-equilibrium equations
governing these variables are:
\begin{eqnarray}
\label{eq:a1}
0 &=& a_1(\Lambda^2+\Omega^2) - 2a_2\Lambda\Omega\nonumber\\
&+&\frac{1}{2}\frac{\wh{V}}{\wh{\mathcal{M}}}R_{NS}^3a_1\tilde{A}_1
+ \frac{R_{NS}^2}{\wh{\mathcal{M}}}\frac{\Pi}{a_1} - c_{11}a_1\\
\label{eq:a2}
0 &=& a_2(\Lambda^2+\Omega^2) - 2a_1\Lambda\Omega\nonumber\\
&+& \frac{1}{2}\frac{\wh{V}}{\wh{\mathcal{M}}}R_{NS}^3a_2\tilde{A}_2
+ \frac{R_{NS}^2}{\wh{\mathcal{M}}}\frac{\Pi}{a_2} - c_{22}a_2\\
\label{eq:a3}
0 &=& \frac{1}{2}\frac{\wh{V}}{\wh{\mathcal{M}}}R_{NS}^3a_3
\tilde{A}_3 + \frac{R_{NS}^2}{\wh{\mathcal{M}}}\frac{\Pi}{a_3} - c_{33}a_3\\
\label{eq:phi=Psi}
\phi &=& \Psi
\end{eqnarray}
where $\wh{V}$ and $\wh{\mathcal{M}}$ are, respectively, the effective
relativistic self-gravity potential and the scalar quadrupole moment
of the NS at spherical equilibrium (see below). $R_{NS}$ is the NS radius and the
$\tilde{A}_i$'s are  defined as
\begin{eqnarray}
\label{def:tildeA}
\tilde{A}_i = \int_0^\infty\frac{d\sigma}{(a_i^2+\sigma)
\sqrt{(a_1^2+\sigma)
    (a_2^2+\sigma)(a_3^2+\sigma)}}\,.
\end{eqnarray}
The $c_{ij}$'s denote the components of the BH tidal tensor in the
principal frame; $\phi$ --- which is connected to $\Omega$ by
$\dot\phi \equiv \Omega$ --- is the angle that brings the
parallel-propagated frame  in the
principal frame, by a rotation around the $a_3$ axis \cite{Marck}. 
Finally, $\Psi$ is an
angle that governs the rotation of the parallel-propagated frame
in order to preserve its parallel transport, and its time evolution is
given by
\begin{eqnarray}
\dot{\Psi} = \sqrt{\frac{M_{BH}}{r^3}}\,,&&\nonumber\\
\end{eqnarray}
where $M_{BH}$ and $r$ are the BH mass and the BH-NS orbital
separation respectively. Notice that the last equation and
Eq.\,(\ref{eq:phi=Psi}) imply that
\begin{eqnarray}
\label{eq:Omega}
\Omega = \sqrt{\frac{M_{BH}}{r^3}}\,.&&
\end{eqnarray}
The fifth equation for the five fluid variables is provided by the
definition of the circulation
\begin{eqnarray}
\mathcal{C} = 
\frac{\wh{\mathcal{M}}}{R_{NS}^2}[(a_1^2+a_2^2)\Lambda-2a_1a_2\Omega]\,,
\nonumber\\
\end{eqnarray}
so that for irrotational fluids one has
\begin{eqnarray}
\label{eq:Lambda}
\Lambda = \frac{2a_1a_2\Omega}{a_1^2+a_2^2}\,.
\end{eqnarray}

The effective
relativistic self-gravity potential is given by
\begin{eqnarray}
\label{def:whV}
\wh{V}=-4\pi \int_0^{R_{NS}}
\frac{d\Phi_{TOV}}{d\hat{r}}\hat{r}^3\hat{\rho} d\hat{r}\,,
\end{eqnarray}
where $d \Phi_{TOV}/d\hat{r}$
is given by the Tolman-Oppenheimer-Volkoff (TOV)
stellar structure equations
\begin{eqnarray}
\label{def:dPhiTOVdr}
\frac{d\Phi_{TOV}}{dr}&=&\frac{[\epsilon(r)+P(r)][m_{TOV}(r)
  +4\pi r^3P(r)]}{\rho(r)
r[r-2m_{TOV}(r)]}\nonumber\\
\label{def:mTOV}
m_{TOV}(r) &=& 4\pi\int_0^r dr'r'^2\epsilon(r')\,.\nonumber
\end{eqnarray}
The scalar quadrupole moment $\wh{\mathcal{M}}$ is defined as
\begin{eqnarray}
\label{def:whM}
\wh{\mathcal{M}} &=& \frac{4\pi}{3}\int_0^{R_{NS}}r^4\rho dr
\end{eqnarray}
and, like $\wh{V}$, must be calculated at spherical
equilibrium. Finally, the relevant Kerr BH tidal tensor components are
\begin{eqnarray}
\label{def:c11}
c_{11} &=& \frac{M_{BH}}{r^3}\left[1-3\frac{r^2+K}{r^2}\cos^2(\Psi-\phi)
\right]\\
c_{22} &=& \frac{M_{BH}}{r^3}\left[1-3\frac{r^2+K}{r^2}\sin^2(\Psi-\phi)
\right]\\
\label{def:c33}
c_{33} &=& \frac{M_{BH}}{r^3}\left( 1+3\frac{K}{r^2}\right)~,
\end{eqnarray}
where 
\begin{eqnarray}
K = (aE-L_z)^2;
\end{eqnarray}
$a$ is the  black hole spin parameter, and  $E$ and $L_z$ are, respectively,
the energy  and  the $z$-orbital angular momentum per unit mass.
Since we consider circular equatorial geodesics, $E$ and $L_z$ are
\begin{eqnarray}
E &=& \frac{r^2-2M_{BH}r+a\sqrt{M_{BH}r}}{r\sqrt{P}}\nonumber\\
L_z &=& \frac{\sqrt{M_{BH}r}(r^2-2a\sqrt{M_{BH}r}+a^2)}{r\sqrt{P}}\,,
\end{eqnarray}
where
\begin{eqnarray}
P = r^2-3M_{BH}r+2a\sqrt{M_{BH}r}\,.
\end{eqnarray}
%

\subsubsection{Numerical integration}\label{compute}
We solve Eqs. (\ref{eq:a1})-(\ref{eq:a3}), with the aid of
Eqs.\,(\ref{eq:Omega}), (\ref{eq:Lambda}) and of definitions
(\ref{def:tildeA}), (\ref{def:whV}), (\ref{def:whM}), 
(\ref{def:c11})-(\ref{def:c33}), by adopting a multidimensional
Newton-Raphson scheme \cite{numrec} in order to determine the values
of the axes of the ellipsoid for each quasi-stationary orbit,
identified by the orbital separation $r$.

We start by solving the TOV stellar structure equations for a
non-rotating spherical neutron star in equilibrium. We then fix the
black hole spin parameter $a$ and the binary mass ratio $q$ and place
the star at a distance $r_0\gg R_{NS}$ from the black hole (we
  obviously make sure that the sequence we obtain is independent of
  $r_0$). Subsequently, we gradually reduce the orbital separation and
solve Eqs.\,(\ref{eq:a1})-(\ref{eq:a3}) at each step and monitor the
star axes until a critical separation $r_{tide}$ is reached,
at which a quasi-equilibrium configuration is no longer possible. This
critical distance physically corresponds to the tidal disruption of
the neutron star. It may be identified by exploiting the fact that the
Newton-Raphson algorithm cannot find any solution to the system of
equations, or by calculating numerically the derivative $\partial
r_{norm}/\partial(a_2/a_1)$, where
\begin{eqnarray}
r_{norm}=\frac{r}{R_{NS}}\left(\frac{M_{NS}}{M_{BH}}\right)^{1/3}\,,\nonumber
\end{eqnarray}
and keeping track of it since it tends to zero at tidal disruption:
both methods yield the same values of $r_{tide}$. We mention that an alternative 
approach to evaluate $r_{tide}$ in a Newtonian framework, based on the 
estimate of the Roche lobe radius, has been used in \cite{LP07}. 
If the tidal disruption is not encountered, the
quasi-equilibrium ends when the neutron star surface crosses the black
hole horizon and hence the coalescence terminates with a plunge.

The quantity $r_{tide}$ has to be compared
with the value of the radius
of the innermost stable circular orbit $r_{ISCO}$, which is determined
by using the formulae derived in \cite{BPT1972} for a point mass in the
gravitational field of a Kerr BH:
\begin{eqnarray}
r_{ISCO} &=& M_{BH}\{3+Z_2\mp [(3-Z_1)(3+Z_1+2Z_2)]^{1/2}\}\nonumber\\
Z_1 &=& 1+(1-a^2/M_{BH}^2)^{1/3}\nonumber\\
&\times&[(1+a/M_{BH})^{1/3}+(1-a/M_{BH})^{1/3}]\nonumber\\
Z_2 &=& (3a^2/M_{BH}^2+Z_1^2)^{1/2}\,,
\label{risco}
\end{eqnarray}
where the upper (lower) sign holds for co- (counter-)rotating
orbits. We remind the reader that if $r_{tide}>
r_{ISCO}$ the star is disrupted before the merger starts, and the
gravitational signal will exhibit a cutoff at a
frequency $\nu_{GWtide}$.

\section{Determination of the cutoff frequency $\nu_{GWtide}$}
\label{nutide}
To compute $\nu_{GWtide}$, we model the inspiral of the mixed
binary by means of a post-Newtonian (PN) approach and truncate the
inspiral when the orbital separation reaches $r_{tide}$; we
then read off the orbital frequency at the tidal disruption
$\nu_{orb~tide}$: this is related to the GW frequency by
$\nu_{GWtide}=2\nu_{orb~tide}$.

We follow a Hamiltonian approach. The conservative part of the
two-body Hamiltonian is known up to order $3$PN,
e.g. \cite{BuonChenDam2006}; however, we shall truncate it at $2$PN
order since  we will use GW dissipation terms of order
$2.5$PN \cite{Schafer1985}. The PN-expanded Hamiltonian for the
relative motion of the BH-NS binary is
\begin{eqnarray}
\label{eq:Horb}
\mathcal{H}_{orb}^{Exp} = \mathcal{H}_{N} +
\mathcal{H}_{PN} + \mathcal{H}_{2PN} +
\mathcal{H}_{SO} + \mathcal{H}_{SS}\,,
\end{eqnarray}
the single contributions being:
\begin{eqnarray}
\label{Eq:H_N}
\mathcal{H}_{N} &=& \frac{1}{2\mu}\left( P_r^2 +
  \frac{P_\varphi^2}{r^2}\right) - \frac{G_{N}\mu
  M_{Tot}}{r} \\
\label{Eq:H_PN}
\mathcal{H}_{PN} &=& \frac{3\eta -1}{8c^2\mu^3}\left( P_r^2 +
  \frac{P_\varphi^2}{r^2}\right)^2\nonumber\\
&-&\frac{G_{N}M_{Tot}}{2c^2\mu r}\left\{(3+\eta)
  \left( P_r^2 + \frac{P_\varphi^2}{r^2}\right) +\eta P_r^2\right\}\nonumber\\
&+&\frac{G_{N}^2\mu M_{Tot}^2}{2c^2r^2}\\
\label{Eq:H_2PN}
\mathcal{H}_{2PN} &=& \frac{1-5\eta+5\eta^2}{16c^4\mu^5}\left( P_r^2 +
  \frac{P_\varphi^2}{r^2}\right)^3\nonumber\\
&+&\frac{G_{N}M_{Tot}}{8c^4\mu^3 r}\left[(5-20\eta
  -3\eta^2)\left( P_r^2 + \frac{P_\varphi^2}{r^2}\right)^2\right.\nonumber\\
&-&\left.2\eta^2P_r^2\left( P_r^2 + \frac{P_\varphi^2}{r^2}\right)
-3\eta^2P_r^4\right]\nonumber\\
&+&\frac{G_{N}^2M_{Tot}^2}{2c^4\mu r^2}
\left[(5+8\eta)\left( P_r^2 + \frac{P_\varphi^2}{r^2}\right)+3\eta P_r^2\right]
\nonumber\\
&-&\frac{G_{N}^3(1+3\eta)\mu M_{Tot}^3}{4c^4r^3}\\
\label{Eq:H_SO}
\mathcal{H}_{SO} &=&\frac{G_{N}}{c^2r^3}{\mathbf L}
\cdot\left(2+\frac{3M_{NS}}{2M_{BH}}\right){\mathbf J}_{BH}\\
\label{Eq:H_SS}
\mathcal{H}_{SS} &=& \frac{G_{N}}{c^2r^3}
[3({\mathbf J}_{BH}\cdot{\mathbf n})({\mathbf J}_{BH}\cdot{\mathbf n})\nonumber\\
&-&({\mathbf J}_{BH}\cdot{\mathbf J}_{BH})]
\frac{M_{NS}}{M_{BH}}
\end{eqnarray}
where $G_N$ is the gravitational constant, $c$ is the speed of light,
$M_{Tot}=M_{BH}+M_{NS}$ is the total mass of the system,
$\mu=M_{BH}M_{NS}/M_{Tot}$ is its reduced mass, $\eta=\mu/M_{Tot}$ is
the symmetric mass ratio, $P_r$ and $P_\varphi$ are the conjugate
variables of the orbital separation $r$ and the orbit angle coordinate
$\varphi$, ${\mathbf L}= {\mathbf r}\times{\mathbf P}$ is the orbital
angular momentum and ${\mathbf J}_{BH}$ is the spin angular momentum
of the black hole. The vectors ${\mathbf L}$ and ${\mathbf J}_{BH}$
are both perpendicular to the orbit plane; the magnitude of the latter
is equal to $aM_{BH}$ and such vector is not evolved as a dynamic
variable in order to follow the spirit of the affine model, according
to which the presence of the NS does not influence the BH. The
spin-orbit (SO) and spin-spin (SS) contributions to
$\mathcal{H}_{orb}^{Exp}$ therefore reduce to
\begin{eqnarray}
\mathcal{H}_{SO} &=&\frac{G_{N}}{c^2r^2}
\sqrt{P_r^2+\frac{P_\varphi^2}{r^2}}\left(2+\frac{3M_{NS}}{2M_{BH}}\right)J_{BH}
\end{eqnarray}
and
\begin{eqnarray}
\mathcal{H}_{SS} &=&\frac{2G_N}{c^2r^3}\frac{M_{NS}}{M_{BH}}J_{BH}^2\,.
\end{eqnarray}
The dynamics described by this Hamiltonian is supplied with GW
dissipation by means of the $2.5$PN non-conservative terms
\begin{eqnarray}
\label{eq:fr}
f_r &=& -\frac{8G_{N}^2}{15c^5\eta r^2}
\left(2P_r^2+\frac{6P_\varphi^2}{r^2}\right)\\
f_\varphi &=& -\frac{8G_{N}^2}{3c^5}\frac{P_rP_\varphi}{\eta r^4}\\
F_r &=& -\frac{8G_{N}^2}{3c^5}\frac{P_r}{r^4}
\left(\frac{P_\varphi^2}{\eta r}-\frac{G_{N}\eta M_{Tot}^3}{5}\right)\\
\label{eq:fvph}
F_\varphi &=& -\frac{8G_{N}^2}{5c^5}\frac{P_\varphi}{\eta r^3}
\left(\frac{2G_N\eta^2M_{Tot}^3}{r} + \frac{2P_\varphi^2}{r^2} - P_r^2\right)
\end{eqnarray}
to be included in the Hamilton equations of motion as
\begin{eqnarray}
\frac{dr}{dt} &=& \frac{\partial\mathcal{H}}{\partial P_r} + f_r\\
\frac{d\varphi}{dt} &=& \frac{\partial\mathcal{H}}{\partial P_\varphi}
+f_\varphi\\
\frac{dP_r}{dt} &=& -\frac{\partial\mathcal{H}}{\partial r} + F_r\\
\frac{dP_\varphi}{dt} &=& -\frac{\partial\mathcal{H}}{\partial\varphi}
+F_\varphi\,.
\end{eqnarray}

Within this Hamiltonian approach, once the binary orbit is evolved
until $r=r_{tide}$, the GW cutoff frequency due to the NS tidal
disruption $\nu_{GWtide}$ is given by
\begin{eqnarray}
\nu_{GWtide}\equiv2\nu_{orb~tide} = \frac{P_\varphi}{\pi\mu r^2}\Bigg|_{r_{tide}}\,.
\end{eqnarray}
%
\section{Equations of state}\label{eos}
The equation of state of matter at densities larger than the saturation density
of nuclear matter, $\rho_0= 2.67 \cdot 10^{14}$ g/cm$^3$, 
is uncertain. At these densities, which are typical of a neutron star core, 
neutrons can no longer be considered as non interacting particles, and
different ways of modeling their interactions lead to 
a different composition, which may include 
heavy baryons, quarks etc..
All available  EOSs of strongly interacting
matter have been obtained within models based on the theoretical knowledge of the
underlying dynamics and constrained, as much as possible, by empirical data.
Modern EOSs are derived within two main, different approaches: non-relativistic nuclear
many-body theory (NMBT) and relativistic mean field theory (RMFT).
As representative of NMBT we choose  two EOSs named APR2 and BBS1.
For APR2 matter consists of neutrons, protons, electrons and muons
($n, p, e, \mu$) in weak equilibrium. The Hamiltonian includes two- and three- nucleon
interaction terms; the two-nucleon  term is the Argonne $v_{18}$ potential
\cite{WSS},  the three-nucleon term is the Urbana IX potential \cite{PPCPW}.
The many-body Schr\"odinger equation is solved using a variational approach
\cite{AP,APR}.
The calculations include relativistic corrections to the two-nucleon
potential, arising from the boost to a frame in which the total momentum of the
interacting pair is non-vanishing. These corrections are necessary to use
phenomenological potentials, describing interactions between nucleons
in their center of mass frame, in a locally inertial frame associated with the
star.  The maximum mass for this EOS is $M_{max}=2.20 M_\odot$.

The matter composition of  BBS1 is the same as in the APR2 model. The EOS is
obtained  using a slightly different Hamiltonian, including
the Argonne $v_{18}$ two-nucleon potential and the Urbana VII 
three-nucleon potential \cite{SPW86}.
The ground state energy is calculated using G-matrix perturbation theory
\cite{BBS200}.
The maximum mass is $M_{max}= 2.01M_\odot$.

As representative of the RMFT approach, we choose the EOSs named 
BGN1H1, GNH3 and BPAL.

The Balberg-Gal (BGN1H1) EOS \cite{BGN1H1} describes matter
consisting of neutrons, protons, electrons, muons and hyperons
($\Sigma$, $\Lambda$ and $\Xi$) in equilibrium. 
The effective potential parameters are tuned in order to 
reproduce the properties of nuclei
and hypernuclei according to high energy experiments.  This EOS is a
generalization of the Lattimer-Swesty EOS \cite{LS}, which does not
include hyperons. The maximum mass is $M_{max}= 1.63~M_\odot$.

The Glendenning (GNH3) EOS \cite{GNH3} considers $n, p, e, \mu$
up to a certain density $\rho_H\simeq
2\rho_0$; beyond this point, additional baryon states (such as the
$\Delta$ and the hyperons $\Lambda, \Sigma, \Xi$) and the
mesons $\pi, \sigma, \rho, \omega, K, K^*$ are introduced. 
Below the hadronization density $\rho_H$ the EOS is very stiff
but causal; the appearance of hyperons strongly softens the EOS
because they are more massive than nucleons and when they start to
fill their Fermi sea they are slow and replace the highest energy
nucleons. The maximum mass is $M_{max}= 1.96~M_\odot$.

The  three  EOSs BPAL12, BPAL22 and BPAL32, are derived using a
density dependent nucleon-nucleon effective interaction 
(as for Skyrme nuclear interactions) \cite{bombaci}.
Matter is composed of $n, p, e, \mu$ in weak equilibrium, no hyperons are present
and the EOS parameters are fixed to reproduce the saturation properties of
nuclear matter.
BPAL12 is  particularly soft, with a nuclear  incompressibility $k =
120$ MeV. Since the empirical value commonly accepted ranges within
$k\in (220-270)$ MeV, it is clear that BPAL12 has to be considered as
an EOS soft extreme, but still compatible with astrophysical observations, since
the maximum mass it predicts is $M_{max}=1.45~M_\odot$. 
BPAL22 and BPAL32 are two different versions of the BPAL EOS, corresponding to
more realistic values of $k$: $k=180$  MeV for BPAL22, with a maximum mass 
$M_{max}=1.72~M_\odot$, and $k=240$  MeV for BPAL32, with a maximum mass 
$M_{max}=1.93~M_\odot$.
At a density of about one half of the nuclear saturation density we match the
EOS of the core with a crust, which is composed of three
layers; the inner layer is the Douchin-Haensel (SLy4) crust
\cite{SLy4a,SLy4b}; for $10^{8}<\rho < 10^{11}\,$g/cm$^3$
and for $\rho < 10^8\,$g/cm$^3$, we use, respectively
the Haensel-Pichon (HP94) EOS \cite{HP94}, and
the  Baym-Pethick-Sutherland (BPS) EOS \cite{BPS}.

\begin{figure}[ht]
\begin{center}
\epsfig{file=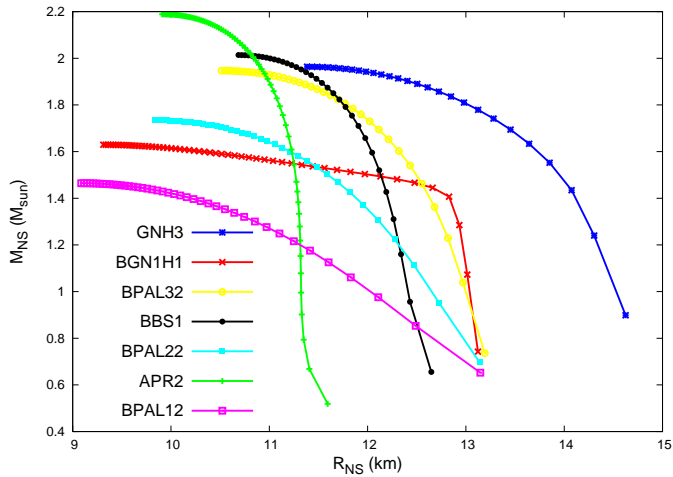,width=180pt,angle=270}
\end{center}
\caption{\it (Color online) The neutron star mass is plotted  versus the stellar
radius for the EOSs used in this paper.}
\label{fig1}
\end{figure}

In Fig.~\ref{fig1} we show the mass-radius diagram for the selected EOSs.
They clearly exhibit very different behaviours.
To some extent,  BPAL12 and GNH3 can be considered, respectively, as
soft and hard extremes. APR2 has a wide mass interval 
where the radius is almost insensitive to mass variations; BGN1H1 also
shows this feature, albeit for a more restricted mass interval,
whereas the BPAL and GNH3 EOSs do not. BGN1H1, on the other hand, has a
sudden softening which does not characterise the other three equations
of state; this softening is due to the appearance of hyperons in the
core of the neutron star above a critical central density.

\section{Error bar on $R_{NS}$ and on $\nu_{GWtide}$}\label{errorbar}
Let us suppose that in a BH-NS coalescence the  neutron star is disrupted
before reaching the ISCO, and that the gravitational wave interferometers detect the
emitted signal which, as discussed in previous sections, has the form of a
chirp truncated at the frequency $\nu_{GWtide}$. 
With  a suitable data analysis, the values of  the symmetric mass ratio
$\eta = M_{BH}M_{NS}/(M_{BH}+M_{NS})^2$ and of the chirp mass $\mathcal{M}=
\eta^{3/5}(M_{BH}+M_{NS})$ can be extracted from the data
with a certain error. These errors have been evaluated for
non-spinning compact binary sources, including up to 3.5PN
terms, in \cite{Arun},\cite{sathiaSchutz}. For instance,
they are displayed in Fig.~12 of \cite{sathiaSchutz}, which shows
the one-sigma fractional errors in $\mathcal{M}$ and $\eta$ for
non-spinning binary black hole sources as a function of the total mass of the
system. These errors reduce when the
dynamical evolution of spins is included, so that they may be regarded
as upper limits. Since a similar analysis has not been performed in
the BH-NS case, in what follows we shall adopt these errors as appropriate
also for mixed binaries. 
In the following we shall consider neutron stars whose mass is in the range
$1.2-1.6$ solar masses and mass ratios from 3 to 15, 
therefore the total mass will range within $\sim 5-26~M_\odot$. In this region the
fractional error on $\mathcal{M}$ is  smaller than $\sim  10^{-3}$
and of the order of  $\sim 1-3 \cdot 10^{-2}$ for $\eta$.
These data refer to advanced LIGO assuming the source at a fixed distance of
$300$ Mpc.

\begin{center}
\begin{figure}[ht]
\epsfig{file=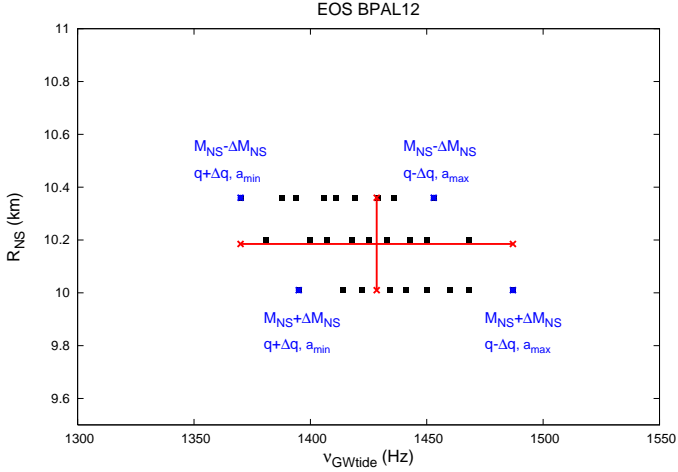,width=180pt,angle=270}
\caption{\it (Color online)  The NS radius is plotted versus $\nu_{GWtide}$
for any possible combination of the following data:
$M_{NS}=1.4\pm 0.02~M_\odot,~
a/M_{BH}=0.5\pm 0.05,~q=5\pm 0.15\,,$ and for the EOS BPAL12.
The horizontal and vertical spread of the data identify the error bars (red
cross).
   }
\label{fig2}
\end{figure}
\end{center}

Since in our case the mass parameters of the binary  are the neutron
star mass and the mass ratio, we express $\mathcal{M}$ and $\eta$ as
\be
\eta = \frac{q}{(1+q)^2},\qquad
\mathcal{M} =\frac{M_{NS}q^{3/5}}{(1+q)^{1/5}}\,.
\label{etamchirp}
\ee
By propagating the errors we find
the following expressions for the one-sigma fractional
errors in the neutron star mass and in the mass ratio:
\beq
\label{Eq:qErr}
\frac{\Delta q}{q} &=& \frac{q+1}{q-1}\frac{\Delta\eta}{\eta}\\
\label{Eq:MNSErr}
\frac{\Delta M_{NS}}{M_{NS}} &=& \frac{\Delta \mathcal{M}}{\mathcal{M}}
+ \frac{2q+3}{5(q-1)}\frac{\Delta \eta}{\eta}\,.
\eeq
The absolute error on $M_{NS}$ and $q$ for the binaries we consider
are given in Table~\ref{table1}.
\begin{table}
\centering
\caption{\it
Absolute errors on the neutron star mass and on the mass ratio.
$\Delta q$ is independent of the neutron star mass.
$\Delta M_{NS}$ is expressed in solar mass units.
}
\vskip 12pt
\begin{tabular}{|c|c|c|c|c|}
\hline
& &$1.2\,M_\odot$ &$1.4\,M_\odot$ &$1.6\,M_\odot$ \\
\hline
$q$&$\Delta q$& $\Delta M_{NS}$ &  $\Delta M_{NS}$   & $\Delta M_{NS}$   \\
\hline
3&0.12&0.02&0.03&0.03\\
5&0.15&0.017&0.02&0.02\\
10&0.24&0.013&0.015&0.018\\
15&0.34&0.012&0.015&0.017\\
\hline
\end{tabular}
\label{table1}
\end{table}
We shall now discuss how the errors
which affect the estimate of the binary parameters influence the evaluation
of the neutron star radius and of $\nu_{GWtide}$.
Since we do not know the error on the black hole angular momentum, in what
follows we shall assume that, with an accurate post-detection data analysis,
$a/M_{BH}$ could be measured with a $10\%$ accuracy.  As an example,
let us suppose that  the neutron star mass and
the mass ratio measured from a detected signal
are, say, $1.4~M_\odot$ and $5$, respectively, plus or minus the
corresponding error which can be found in Table~\ref{table1}.
In Fig.~\ref{fig2} we plot the neutron star radius versus $\nu_{GWtide}$,
evaluated as explained in section \ref{nutide}, for any possible combination
of the following data:
\be
M_{NS}=(1.4\pm 0.02)~M_\odot,~
\frac{a}{M_{BH}}=0.5\pm 0.05,~q=5\pm 0.15\,,
\label{errors}
\ee
and for a given EOS, for instance BPAL12.
The data are spread in a region which can be identified by a vertical 
and a horizontal error bar, indicated as a red cross in the figure: the 
vertical error corresponds to the maximum and minimum values of the neutron
star mass.  
The horizontal error is identified by the points 
$(M_{NS}-\Delta M_{NS}, q+\Delta q, a_{min})$ and 
$(M_{NS}+\Delta M_{NS}, q-\Delta q, a_{max})$.  
In the following figures, where we will show the graphs 
of $R_{NS}$ versus $\nu_{GWtide}$
for different values of the binary parameters and for different EOSs, 
we shall plot directly the cross, determined as in Fig.~\ref{fig2}.

\section{Results}\label{results}
\begin{center}
\begin{figure}[ht]
\epsfig{file=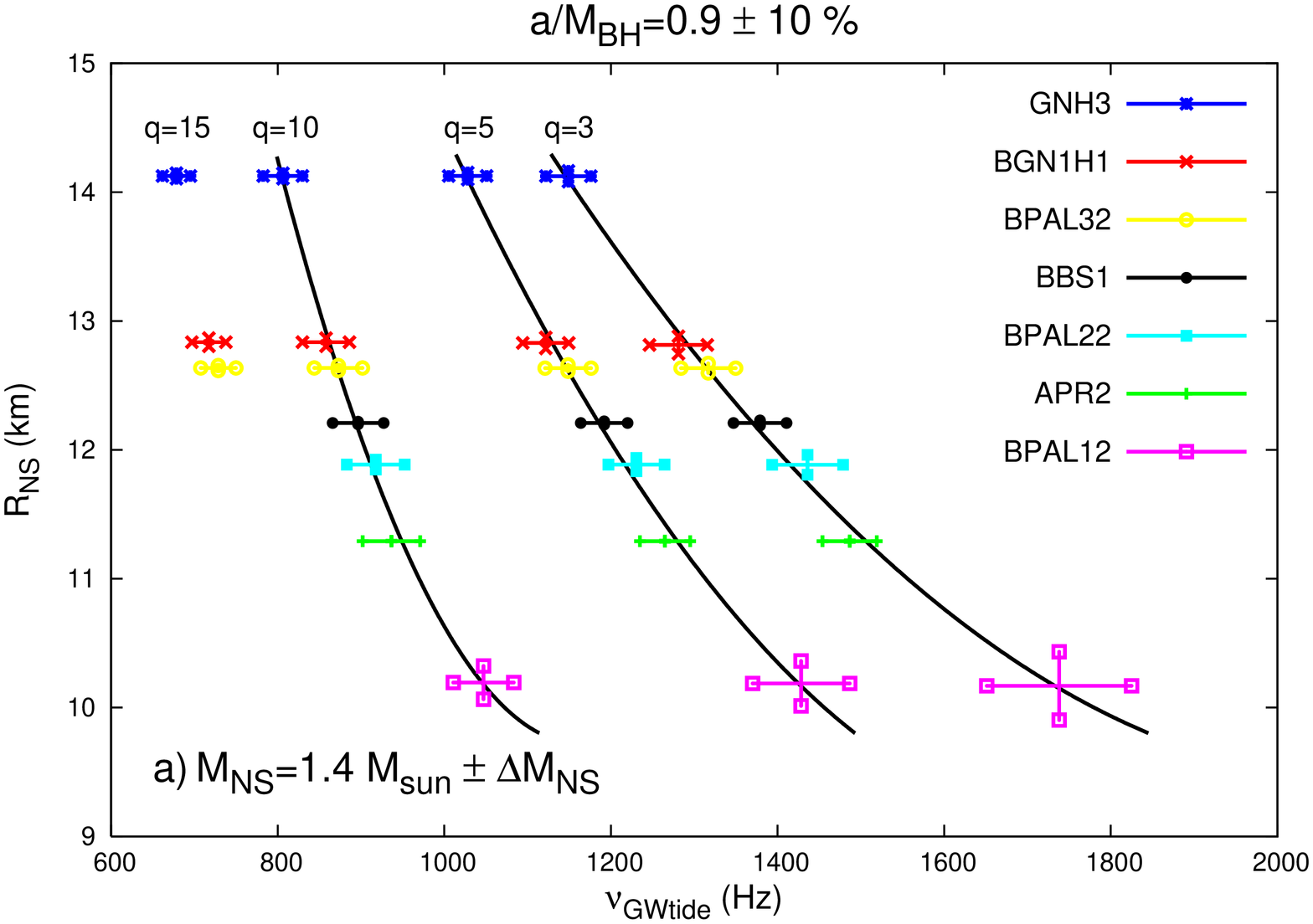,width=180pt,angle=270}
\epsfig{file=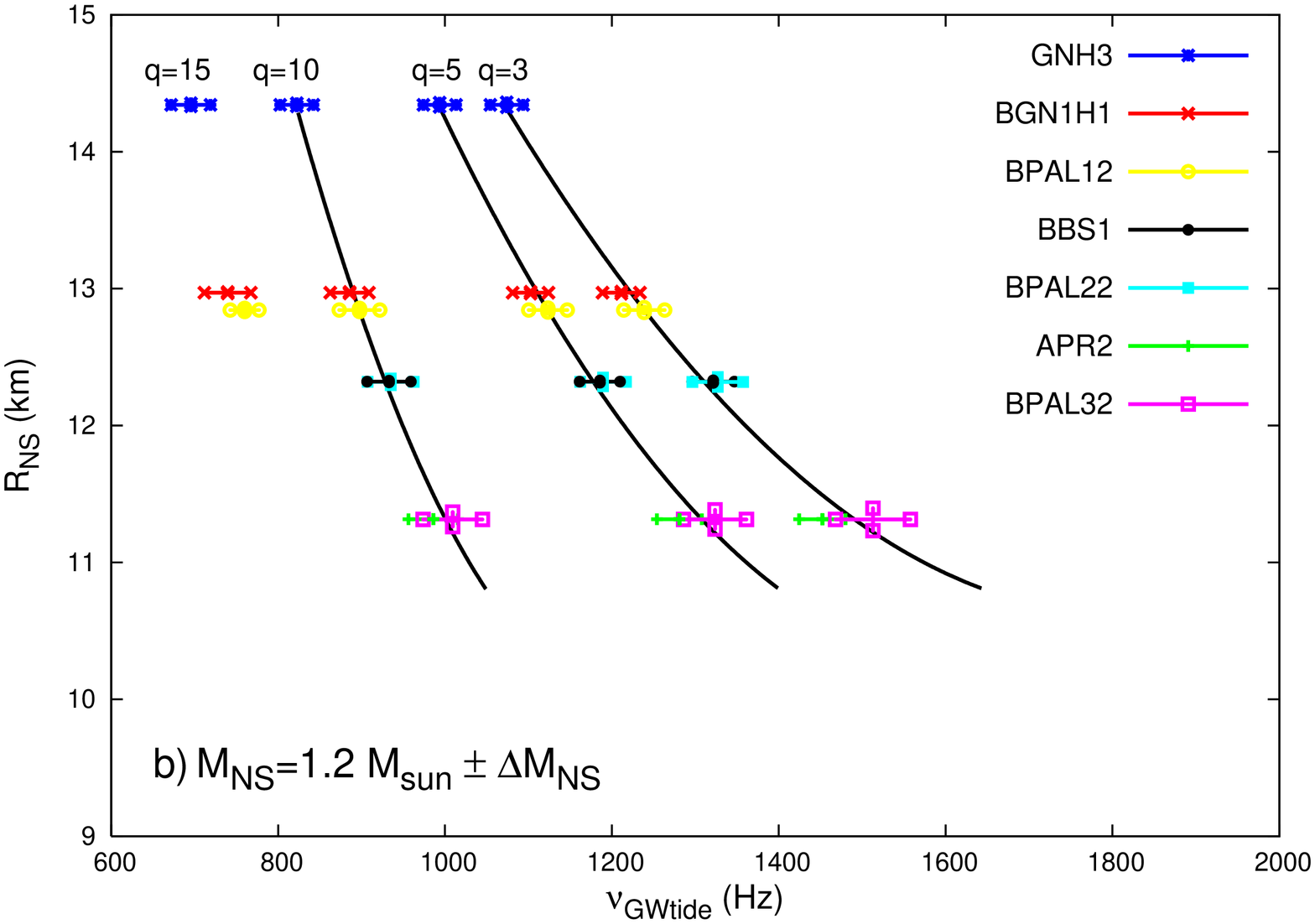,width=180pt,angle=270}
\caption{\it (Color online)  $R_{NS}$ is plotted versus 
$\nu_{GWtide}$, for  a black hole with angular momentum
${a}/{M_{BH}}=0.9\pm 0.09$ and a neutron star with
mass $M_{NS}=1.4~ M_\odot\pm \Delta M_{NS}$ (upper panel) and 
$M_{NS}=1.2~ M_\odot\pm \Delta M_{NS}$ (lower panel).
The data are plotted for different values of the mass ratio $q\pm \Delta q$ 
and for the EOSs considered in section~\ref{eos}.
The values of $\Delta M_{NS}$ and $\Delta q$ are given in Table~\ref{table1}.
Black continuous lines are the parabolic fits of the data corresponding to
each EOS at fixed $q$.
  }
\label{fig3}
\end{figure}
\end{center}

\begin{center}
\begin{figure}[ht]
\epsfig{file=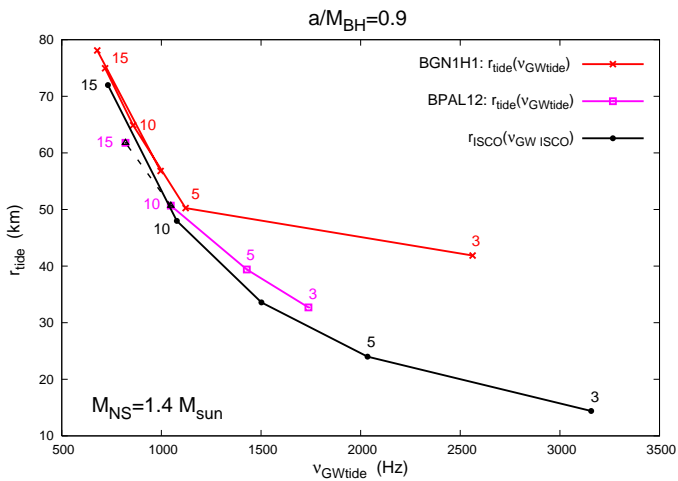,width=180pt,angle=270}
\caption{\it (Color online)  The tidal disruption radius, $r_{tide}$, is plotted versus 
$\nu_{GWtide}$ for the EOSs BGN1H1 (upper curve) and  BPAL12 (middle curve), 
assuming $M_{NS}=1.4~M_\odot$ and $a/M_{BH}=0.9$. The numbers label
points corresponding to different values of $q$. 
For the same values of $q$ we also plot the value of $r_{ISCO}$ (lower curve)
versus the corresponding frequency $\nu_{GW~ISCO}$ (see text).
The dashed part of the curve for BPAL12 refers to values of $q$ for which 
the star is not disrupted before the ISCO, i.e. for which $r_{tide} <
r_{ISCO}$. 
 }
\label{fig4}
\end{figure}
\end{center}

\begin{center}
\begin{figure}[ht]
\epsfig{file=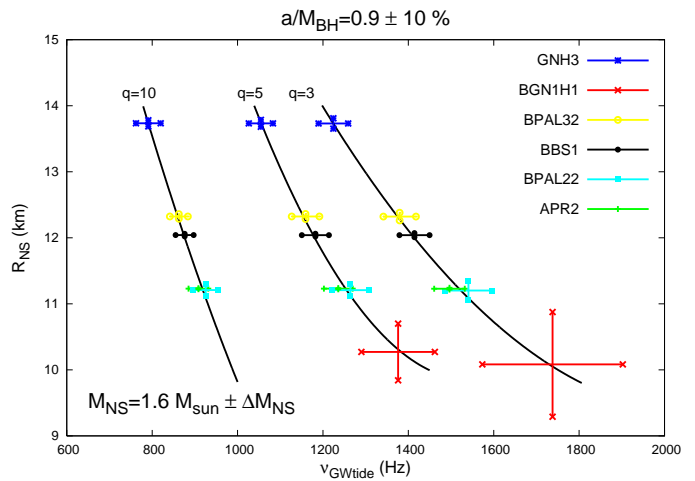,width=180pt,angle=270}
\caption{\it (Color online)  The same plot as in Fig.~\ref{fig3}, but for 
$M_{NS}= 1.6 M_\odot\pm \Delta M_{NS}$.
 }
\label{fig5}
\end{figure}
\end{center}

\begin{center}
\begin{figure}[ht]
\epsfig{file=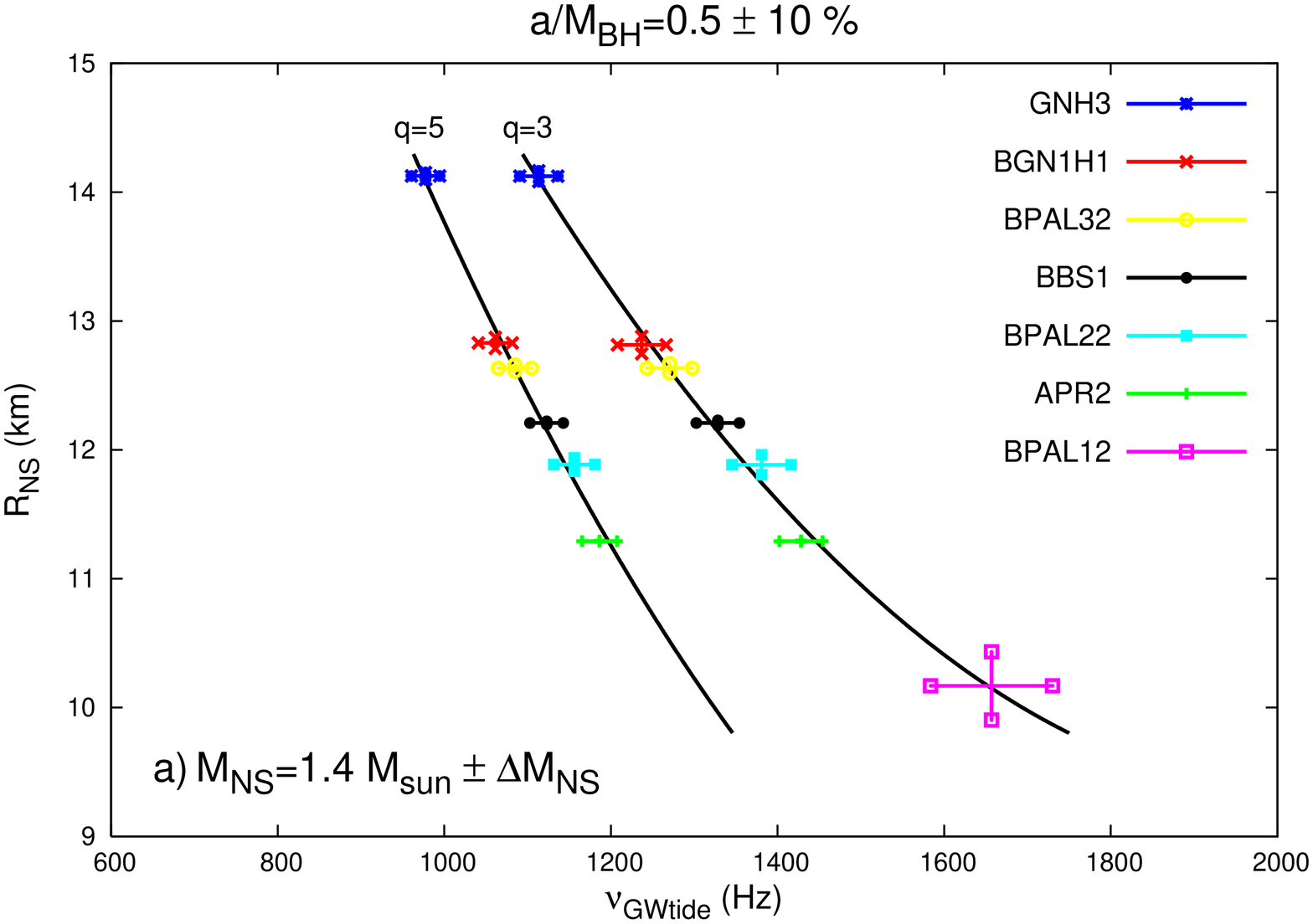,width=180pt,angle=270}
\epsfig{file=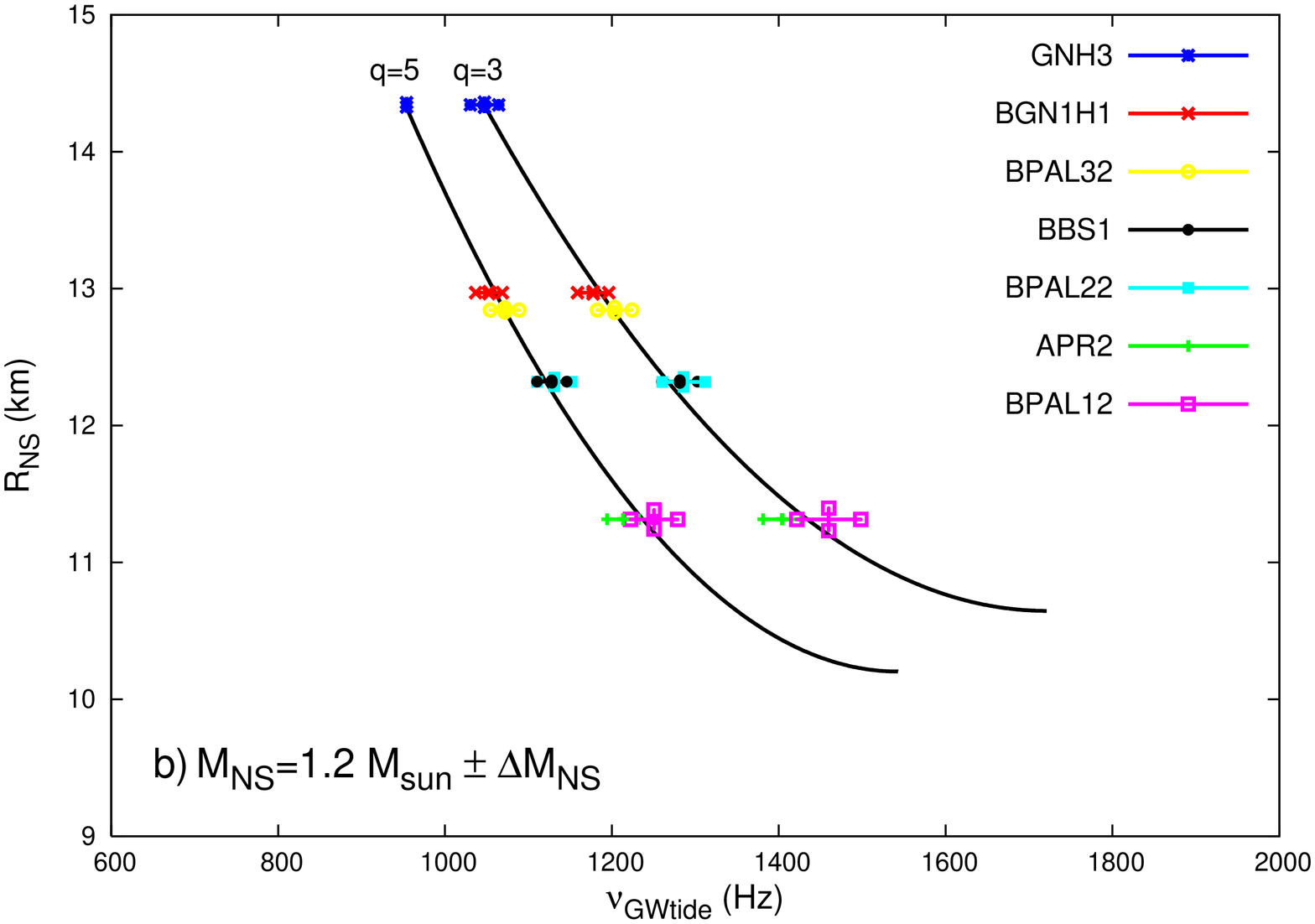,width=180pt,angle=270}
\caption{\it (Color online)  The same plot as in Fig.~\ref{fig3} for a black hole with
angular momentum ${a}/{M_{BH}}=0.5\pm 0.05$. The data for  $q=10$ and $15$ are
missing because no disruption occurs
for the considered EOSs.}
\label{fig6}
\end{figure}
\end{center}

\begin{center}
\begin{figure}[ht]
\epsfig{file=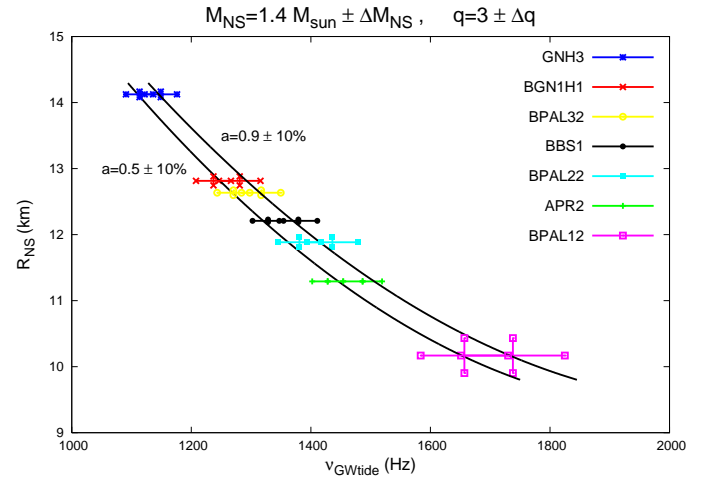,width=180pt,angle=270}
\caption{\it (Color online)  The  plot of  $R_{NS}$ versus $\nu_{GWtide}$ 
is done for the same NS mass,  $M_{NS}=1.4~M_\odot \pm \Delta M_{NS}$,
the same mass ratio $q=3\pm\Delta q$, and
two different values of the angular momentum, $a=0.9\pm 10 \%$ and
$a=0.5\pm 10 \%$.
}
\label{fig7}
\end{figure}
\end{center}

\begin{center}
\begin{figure}[ht]
\epsfig{file=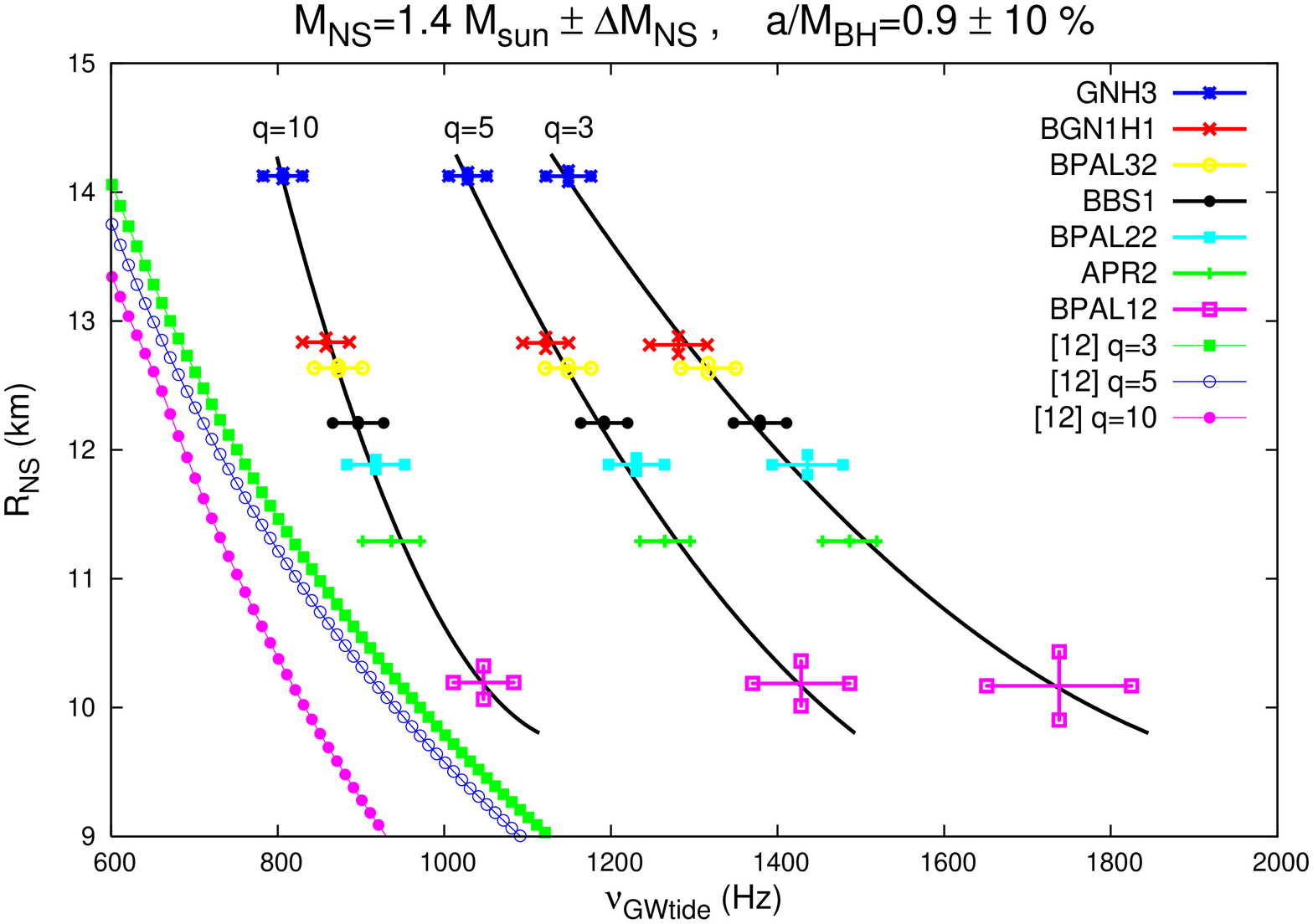,width=180pt,angle=270}
\caption{\it (Color online)  The data of Fig.~\ref{fig3} are compared to the fit
given in Eq. \ref{vallisn} \cite{vallisneri}.   }
\label{fig8}
\end{figure}
\end{center}

\begin{center}
\begin{figure}[ht]
\epsfig{file=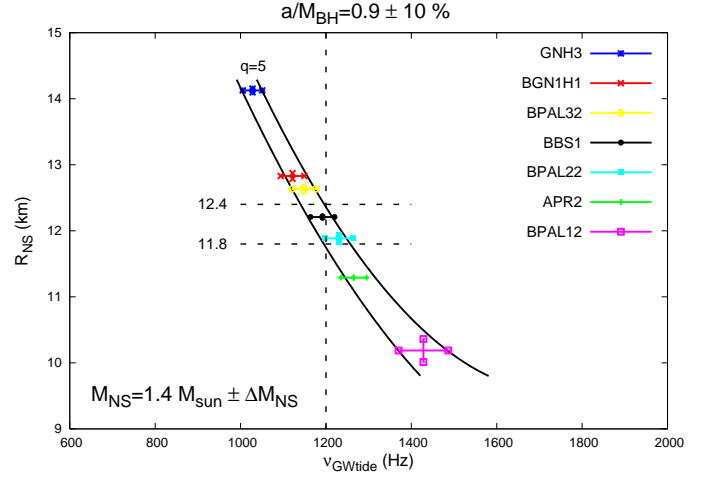,width=180pt,angle=270}
\caption{\it  (Color online)   $R_{NS}$  versus $\nu_{GWtide}$ is plotted
for $M_{NS}=1.4 ~ M_\odot\pm \Delta M_{NS}$, $q=5 \pm \Delta q$ and 
${a}/{M_{BH}}=0.9\pm 0.09$. The black continuous lines are the parabolic fits
of the extreme points of  the horizontal error bar for each EOS. 
The vertical dashed line corresponds to the supposed
detected value of $\nu_{GWtide}$. The two horizontal dashed lines identify the
uncertainty on the evaluation of $R_{NS}$.
} 
\label{fig9}
\end{figure}
\end{center}

The results of our calculations are summarised in the Figs.~\ref{fig3}-\ref{fig9}.
In Fig.~\ref{fig3} we plot the NS radius, $R_{NS}$, versus the cutoff frequency,
$\nu_{GWtide}$, for  a black hole with angular momentum
${a}/{M_{BH}}=0.9\pm 0.09$, and for
a neutron star with mass
$M_{NS}=1.4 ~ M_\odot\pm \Delta M_{NS}$ in Fig.~\ref{fig3}(a), and
$M_{NS}=1.2 ~ M_\odot\pm \Delta M_{NS}$  in
Fig.~\ref{fig3}(b).
The data are plotted for different values of the mass ratio $q\pm \Delta q$
(we omit writing ``$\pm \Delta q$'' in the figures)
and for the considered EOSs.
The values of $\Delta M_{NS}$ and $\Delta q$ 
 are given in Table~\ref{table1}.
For each value of $q$,
the continuous black lines are parabolic fits of the data corresponding to
each EOS.
Fig.~\ref{fig3} shows that, for a given EOS, as $q$ increases
$\nu_{GWtide}$ decreases. Moreover, 
for $q=15$ the data corresponding to the EOSs BBS1, BPAL22, APR2, BPAL12
are missing.
This behaviour is easily understood if we plot the radius 
at which disruption occurs, $r_{tide}$, versus $\nu_{GWtide}$  for different
values of $q$.
For instance, in Fig.~\ref{fig4} this plot is done for 
$M_{NS}=1.4~M_\odot$ and $a/M_{BH}=0.9$ for the EOSs BGN1H1 and
BPAL12.
For comparison, we also plot the value of $r_{ISCO}$, given by Eqs.
(\ref{risco}), 
versus the frequency $\nu_{GW~ISCO}$ of the gravitational signal  emitted when
a point mass of mass $M=M_{NS}$, orbiting  around a Kerr black hole of
mass $M_{BH}=qM$ and angular momentum $a/M_{BH}=0.9$,  reaches  $r_{ISCO}$. 
For  the BGN1H1 star (upper curve) we see that 
$r_{tide}$ is larger than $r_{ISCO}$ for all values of $q$;
thus the  GW signal emitted by these systems 
will exhibit a frequency cutoff at
$\nu_{GWtide}$. Moreover, as $q$ increases the value of
$r_{tide}$ increases, i.e. the star is disrupted at larger distances
from the black hole. As a consequence $\nu_{GWtide}$ is a decreasing function
of $q$, as shown in Fig.~\ref{fig3} for all EOSs.
A similar behaviour is shown by the EOS BPAL12 (middle curve) up to $q=10$.
At that point the curve $r_{tide}$-versus-$\nu_{GWtide}$ crosses 
the lower curve $r_{ISCO}$-versus-$\nu_{GW~ISCO}$, and for larger $q$
$r_{tide}$  becomes smaller than $r_{ISCO}$; this means that, for $q>10$,
the BPAL12 star would be disrupted after reaching the ISCO and the emitted
signal would not exhibit a frequency cutoff. For this reason the part of curve
between the points corresponding to $q=10$ and $q=15$ is indicated as a dashed line.
Going back to Fig.~\ref{fig3}, the above discussion clarifies why for
$q=15$ the data for the EOSs BBS1, BPAL22, APR2 and BPAL12 are missing: the
star merges with the black hole  without being disrupted,
and there is no frequency cutoff in the emitted GW signal.

In Fig.~\ref{fig3} we also notice that the 
vertical errors are always much smaller than the horizontal ones, 
except for the EOS BPAL12, when $M_{NS}=1.4 M_\odot\pm \Delta M_{NS}$.
The reason is that the vertical error is due to the error on the
neutron star mass, which we recall is of the order of a few percent;
the curves corresponding to the different EOSs in the mass-radius diagram of
Fig.~\ref{fig1} show that, for $M_{NS}=1.4~M_\odot$ or
$M_{NS}=1.2~M_\odot$, the stellar radius does not change significantly for
such a small change in $M_{NS}$.
However,  $M_{NS}=1.4~M_\odot$ is close to the maximum mass of  the BPAL12 EOS,
and in its neighborhood the mass-radius curve
is almost flat.
For this EOS even a small change in the mass corresponds
to a significant change in the radius.

Fig.~\ref{fig3}(b) shows that, when $M_{NS}=
1.2 M_\odot\pm \Delta M_{NS}$, for the pairs of EOSs (BBS1,BPAL22) and
(APR2,BPAL12) the data nearly coincide. The reason is that, as shown in
Fig.~\ref{fig1}, for $M_{NS}=1.2~M_\odot$ the stars corresponding to these
pairs of EOSs have nearly the same radius. For the same reason the values of
$\nu_{GWtide}$ for  BGN1H1 and BPAL32 are very close.
In Fig.~\ref{fig5}  the same quantities of Fig.~\ref{fig3} are plotted for   
a larger mass, $M_{NS}= 1.6 M_\odot\pm \Delta M_{NS}$.
The data for the EOS BPAL12 are missing because the maximum mass allowed 
by this EOS is  $M_{max}=1.45~M_\odot$.
A comparison with Fig.~\ref{fig3}(a) shows that, 
for $M_{NS}= 1.6 M_\odot$, the data for
the EOS BGN1H1  move at the bottom of the figure, i.e. for these masses
the NS radius is
smaller than that given by the other EOSs, as can also be seen  from
Fig.~\ref{fig1}; moreover,  the vertical error bars are larger,
because for this EOS $M_{NS}= 1.6 M_\odot$ is very close
to the maximum mass, where the mass-radius curve is nearly flat (see
Fig.~\ref{fig1}). 
In Fig.~\ref{fig6} the same quantities of Fig.~\ref{fig3} 
are plotted   for a black hole with angular momentum
${a}/{M_{BH}}=0.5\pm 0.05$.  Fig.~\ref{fig6}(a) 
refers to a neutron star with  mass $M_{NS}=1.4~M_\odot \pm \Delta
M_{NS}$, whereas in Fig.~\ref{fig6}(b)  $M_{NS}=1.2~M_\odot \pm \Delta
M_{NS}$.
The data for $q=10$ and $15$ are missing because no disruption occurs 
for the considered EOS: tidal disruption is favoured by large
values of the black hole angular momentum.

If  we compare  this figure with Fig.~\ref{fig3}
we see that, for a given mass and EOS, the value of the cutoff frequency
is smaller if the black hole angular momentum  decreases. This means that the tidal
disruption radius $r_{tide}$ increases as $a$ decreases.
It should be noted that for a fixed value of $M_{NS}$ and of $q$
$r_{ISCO}$ also increases as $a$ decreases; it increases faster
than $r_{tide}$, therefore there exists a critical value of the black hole
angular momentum {\it below} which the star  is not disrupted before the ISCO.
In a similar way, for  fixed $a$, $M_{NS}$ and EOS, there exists a critical
value of $q$ {\it above} which no disruption occurs. 
\begin{table}
\centering
\caption{\it
The values of $q$ above which no disruption occurs, $q_{max}$, and the
corresponding
minimum value of $\nu_{GWtide}$, $\nu_{min}$, are tabulated
for the binaries considered in Figs.~\ref{fig3} and \ref{fig6}, i.e for a black hole with
${a}/{M_{BH}}=0.5, 0.9$ and a neutron star with mass $M=1.2, 1.4\,M_\odot$.
}
\vskip 12pt
\begin{tabular}{|c|c|c|c|c|}
\hline
\multicolumn{5}{|c|}{$a/M_{BH}=0.5$}\\
\hline
EOS &\multicolumn{2}{|c|}{$1.2\,M_\odot$} &\multicolumn{2}{|c|}{$1.4\,M_\odot$} \\
\hline
& $q_{max}$ & $\nu_{min}$ (Hz) & $q_{max}$ & $\nu_{min}$ (Hz) \\
\hline
GNH3 & $9.4$ &$780$& $7.5$ &$840$\\
BGN1H1 & $8.4$ &$878$& $6.7$ &$942$\\
BPAL32 & $8.2$ &$899$& $6.6$ &$967$\\
BBS1 & $7.6$ &$966$& $6.1$ &$1032$\\
BPAL22 & $7.6$ &$968$& $5.8$ &$1084$\\
APR2 & $6.9$ &$1068$& $5.6$ &$1128$\\
BPAL12 & $6.6$ &$1116$& $4.6$ &$1375$\\
\hline
\multicolumn{5}{|c|}{$a/M_{BH}=0.9$}\\
\hline 
GNH3 & $23.2$ &$559$& $18.5$ &$596$\\ 
BGN1H1 & $20.6$ &$628$& $16.5$ &$667$\\ 
BPAL32 & $20.1$ &$642$& $15.9$ &$690$\\ 
BBS1 & $18.8$ &$686$& $15.1$ &$726$\\ 
BPAL22 & $18.9$ &$689$& $14.4$ &$759$\\ 
APR2 & $17.0$ &$757$& $13.8$ &$793$\\ 
BPAL12 & $16.2$ &$790$& $11.3$ &$960$\\
\hline
\end{tabular}
\label{table2}
\end{table}
As an example, in
Table~\ref{table2} we give the values of $q_{max}$ and the corresponding
$\nu_{GWtide}$, which is the minimum value of the cutoff frequency to be
expected, for the binaries considered in Figs.~\ref{fig3} and \ref{fig6}.

A direct comparison of the effect of the angular momentum on $\nu_{GWtide}$ 
is displayed in Fig.~\ref{fig7} where the  plot  of $R_{NS}$ versus $\nu_{GWtide}$ is
done for $M_{NS}=1.4~M_\odot \pm \Delta M_{NS}$, $q=3\pm\Delta q$
and for $a/M_{BH}=0.9\pm 10 \%$ and $0.5\pm 10 \%$.
We see that the effect is small, but not negligible.

An interesting feature common to the  Figs.~\ref{fig3}, \ref{fig5} and \ref{fig6}
is that, for the considered EOSs, all data are well fitted by a parabolic fit. 
It should be noted that the fit proposed by Vallisneri in ref.
\cite{vallisneri}, i.e.
\be
\frac{R_{NS}}{M_{NS}^{1/3}M_{BH}^{2/3}}\approx 
\left\{ \begin{array}{rr} 
0.145(\tilde{\nu}M_{BH})^{-0.71} &\hbox{for}~ \tilde{\nu}M_{BH} \lesssim
0.045\\
0.069(\tilde{\nu}M_{BH})^{-0.95} &\hbox{for}~ \tilde{\nu}M_{BH} \gtrsim 
0.045\\
\end{array}\right.
\label{vallisn}
\ee
(where $\tilde{\nu}\equiv\nu_{GWtide}$) 
for $M_{NS}=1.4~M_\odot$ and $M_{BH} =(2.5-80)~M_\odot$, predicts values of 
$R_{NS}$ and $\nu_{GWtide}$ largely 
different from those we find, as shown in Fig.~\ref{fig8}.
In particular we see that, for a given value of the NS mass,  of the radius
(i.e. of the stellar compactness) and of the mass ratio,
the value of $\nu_{GWtide}$ evaluated in \cite{vallisneri} is systematically
smaller than the value we find. This happens for a number of reasons.
For instance, let us consider two binary systems having a black hole of the
same mass  and a neutron star with the same mass and
compactness; in one case the NS structure is computed using the equations of
Newtonian gravity and an $n=1$ polytropic  EOS, in the other case
 using the TOV equations and one of the EOS we 
consider in this paper. If we evaluate the value of $r_{tide}$ for the first
system using the affine approach in Newtonian gravity as in \cite{vallisneri}, 
and for the second system using our improved affine approach, we
will always find  $r_{tide~Newtonian} > r_{tide~improved}$. As a consequence 
the value of $\nu_{GWtide}$ evaluated by the Newtonian approach will be
smaller than that evaluated  with our improved approach.
Furthermore, given the value of $r_{tide}$, in \cite{vallisneri}
$\nu_{GWtide}$ is calculated using
the formula for a point mass moving on a circular orbit in Kerr spacetime,
while we compute this quantity  using the 2.5 Post Newtonian equations
describing the orbital evolution of a binary system. This introduces a further
difference which makes our $\nu_{GWtide}$ larger than that evaluated with the
geodesic approximation, and the difference increases when we consider small 
values of the mass ratio $q$, as shown in Fig.~\ref{fig8}.

The parameters $\alpha,\beta,\gamma$ of our parabolic fits
\be
R_{NS}=\alpha+\beta\nu_{GWtide}+\gamma\nu_{GWtide}^2~,
\ee 
where $R_{NS}$ is expressed in km and $\nu_{GWtide}$ in Hz, are given in
Table~\ref{table3}.
\begin{table}
\centering
\caption{\it
The parameters $\alpha,\beta,\gamma$ of our parabolic fits.
$\alpha$ is in km, $\beta$ in km/Hz, $\gamma$ in km/Hz$^2$.}

\begin{tabular}{|c|c|c|c|c|c|c|}
\hline
\multicolumn{7}{|c|}{$a/M_{BH}=0.5$}\\
\hline
&\multicolumn{3}{|c|}{$1.2\,M_\odot$}
&\multicolumn{3}{|c|}{$1.4\,M_\odot$} \\
\hline
$q$ & $\alpha$ & $\beta$ & $\gamma$ & $\alpha$ & $\beta$ & $\gamma$  \\
\hline
3& 34.7&-2.8$\cdot10^{-2}$&8.1$\cdot10^{-6}$
&32.4&-2.3$\cdot10^{-2}$&5.5$\cdot10^{-6}$
\\
5&38.5&-3.6$\cdot10^{-2}$&11.9$\cdot10^{-6}$
&35.3&-2.9$\cdot10^{-2}$&7.5$\cdot10^{-6}$
\\
\hline
\multicolumn{7}{|c|}{$a/M_{BH}=0.9$}\\
\hline
3& 33.2&-2.5$\cdot10^{-2}$&6.9$\cdot10^{-6}$
&31.7&-2.1$\cdot10^{-2}$&5.0$\cdot10^{-6}$\\ 
5& 37.1&-3.3$\cdot10^{-2}$&10.2$\cdot10^{-6}$
&37.4&-3.2$\cdot10^{-2}$&9.0$\cdot10^{-6}$\\ 
10& 50.0&-6.7$\cdot10^{-2}$&27.7$\cdot10^{-6}$
&56.3&-8.0$\cdot10^{-2}$&34.5$\cdot10^{-6}$\\
15& 47.9&-7.1$\cdot10^{-2}$&32.4$\cdot10^{-6}$
&189.8&-0.47&31.4$\cdot10^{-5}$\\
\hline
\end{tabular}
\label{table3}
\end{table}

We shall now discuss how the results of this paper could
be used to estimate the radius of the star and to gain information on
the equation of state, using gravitational wave detection.
Let us assume that a gravitational wave signal has been detected
and exhibits a frequency cutoff at, say, $\nu_{GWtide}=1200$ Hz. 
A suitable data analysis on the chirp part of the signal allows to determine
the neutron star mass, the mass ratio, with the corresponding uncertainty
given in Table~\ref{table1}, and the black hole angular momentum
with, say, an error of about 10\%. Let us suppose that their values are
$M_{NS}=1.4~ M_\odot\pm \Delta M_{NS}$, $q=5 \pm \Delta q$ and 
${a}/{M_{BH}}=0.9\pm 0.09$, with $\Delta M_{NS}$ and $\Delta q$ given in
Table~\ref{table1}.
With this information, we compute the NS radius versus $\nu_{GWtide}$ for
the considered EOSs as explained in section \ref{compute}, and plot the results 
in Fig.~\ref{fig9}. The two continuous black lines are the parabolic fits
of the points which, for each EOS, are at the extreme of  the
horizontal error bar.
If we draw a vertical line corresponding to the supposed detected cutoff
frequency, i.e.  $\nu_{GWtide}=1200$ Hz, we see that it intercepts the two
lines in two points, which identify a region within which
the stellar radius  should fall.
The data shown in Fig.~\ref{fig9} tell us that we would determine the 
NS radius with an error of 2.5\%. The error would be smaller if the 
detected value of $\nu_{GWtide}$ is  smaller, and larger if it is  larger.
For instance for $\nu_{GWtide}=1400$  Hz it would be 3.3\%,
and 2.2\% for $\nu_{GWtide}=1100$ Hz.

Moreover, we would be able to 
exclude the EOSs which fall outside the region framed by the 
horizontal dashed lines, putting  strict constraints on the 
equation of state inside the neutron star.

\section{Concluding remarks}\label{concl}
In this paper we show that, by detecting a gravitational wave signal emitted
by a BH-NS coalescing binary which exhibits a frequency cutoff due to the
disruption of the star before the ISCO, we may be able to determine the radius
of the star with  quite a good accuracy, of the order of a few percent,
 and to put strict constraints on
the equation of state of matter in the neutron star core. 

Our study does not intend to be exhaustive, since  many more equations of state
may be considered in the analysis; for instance we did not consider 
quark stars. There may exist other branches in the plots 
shown in Figs.~\ref{fig3}-\ref{fig7} corresponding to more exotic EOSs. 

However, from our study it emerges that the quantity which mostly 
affects $\nu_{GWtide}$ is the stellar compactness;
therefore, we expect that in general the cutoff frequencies lay on 
the parabolic fits which correspond to a given neutron star mass, 
mass ratio and  black hole angular momentum.

A further point which should be discussed is the following. In our analysis
we have assumed that  the value of $\nu_{GWtide}$ is known from the detection
of a gravitational wave signal,  but
of course  this quantity also is affected by uncertainties. 
For instance, we do not know how quickly the amplitude of 
the chirp goes to zero at tidal disruption, and therefore  
how sharp the step in the Fourier transform 
of the gravitational signal $h(\nu)$ at $\nu_{GWtide}$ is.
To answer this question, an
accurate modeling of the BH-NS coalescence process is certainly needed.
Moreover, assuming a given ``slope'' in  $h(\nu)$ at  tidal disruption,
how large would the experimental error in the determination of
$\nu_{GWtide}$  be? 
We plan to investigate this problem with a suitable data analysis study on the 
data of Virgo.

\section*{Acknowledgements}
We are indebted to  Ignazio Bombaci for kindly providing the data for the
EOSs BPAL12, 22 and 32, and to Omar Benhar for useful suggestions and discussions.

This work was partially supported by CompStar, a research networking
program of the European Science Foundation.  L.G. has been partially
supported by Grant No. PTDC/FIS/098025/2008. F.P. was supported in
part by DFG Grant SFB/Transergio7 ''Gravitational Wave Astronomy''.


\end{document}